\documentclass[twocolumn,amsmath,amssymb]{revtex4}
\usepackage{graphicx}
\usepackage{dcolumn}
\usepackage{bm}
\begin{document}

\title{Interaction of graphene monolayer with ultrashort laser pulse}

\author{Hamed Koochaki Kelardeh$^1$}
\author{Vadym Apalkov$^1$}
\author{Mark I. Stockman$^{1,2,3}$}
\affiliation{
$^1$Department of Physics and Astronomy, Georgia State
University, Atlanta, Georgia 30303, USA\\
$^2$Fakult\"at f\"ur Physik, Ludwig-Maximilians-Universit\"at, Geschwister-Scholl-Platz 1, D-80539 M\"unchen, Germany\\
$^3$Max-Planck-Institut f\"ur Quantenoptik, Hans-Kopfermann-Strasse 1, D-85748 Garching, Germany
}

\date{\today}
\begin{abstract}
We study the interaction of graphene with ultrashort few femtosecond long optical pulse. For such a short pulse, the electron dynamics is coherent and is described within the tight-binding model of graphene.  The interaction of optical pulse with graphene is determined by strong wave vector dependence of the interband dipole matrix elements, which are singular at the Dirac points of graphene.  The electron dynamics in optical pulse is highly 
irreversible with large residual population of the conduction band. The residual conduction band population as a function of the wave vector is nonuniform with a few localized spots of high conduction band population. The spots are located near the Dirac points and the number of spots depends on the pulse intensity. The optical pulse propagating through graphene layer generates finite transferred 
charge, which, as a function of pulse intensity, changes its sign. At small pulse intensity, the charge is transferred in the direction of the pulse maximum,  
while at large pulse intensity, the direction of the
charge transfer is opposite to the direction of pulse maximum. This property opens unique possibility 
of controlling the direction of the charge transfer by variation of the pulse 
intensity.  
\end{abstract}
\maketitle

\section{Introduction}

Interaction of ultrashort and strong optical laser pulse with solids has been a subject of intensive 
theoretical and experimental research during the last few decades.
\cite{Corkum_et_all_JOPB_Attosecond_Ionization_SiO2, Baltuska_et_al_Attosecond_IonizationPRL_2011,Krausz_et_al_PRL_1998_Femtosecond_Breakdown_Dielectrics, 
Fowler_Nordheim_Proc_Royal_Soc_London_1928_Electron_Field_Emission,  Zener_Proc_Royal_Soc_1934_Breakdown,  Keldysh_JETP_1957_Zener_Ionization,  Wannier_PR_1960_Wannier_States_in_Strong_Fields, Lenzlinger_Snow_J_Appl_Phys_1969_Fowler_Nordheim_Experiment, Hommelhoff_et_al_Nature_2011_As_Tip_Electron_Emission, Reis_et_al_Nature_Phys_2011_HHG_from_ZnO_Crystal, Stockman_et_al_PRL_2010_Metallization, Murnane_et_al_PRL_97_113604_2006_Laser_Assisted_Photoelectric_Effect_from_Surfaces, Stockman_et_al_PRL_2011_Dynamic_Metallization,Schiffrin_at_al_Nature_2012_Current_in_Dielectric,Schultze_et_al_Nature_2012_Controlling_Dielectrics} The interest in this field has grown 
after experimental realization of short laser pulses with just a few oscillations of optical field, which is  comparable to the internal fields of a solid.\cite{Corkum_et_all_JOPB_Attosecond_Ionization_SiO2, Baltuska_et_al_Attosecond_IonizationPRL_2011,Krausz_et_al_PRL_1998_Femtosecond_Breakdown_Dielectrics} Such high intensity optical pulses strongly affect the electron 
dynamics and strongly modify 
the transport and optical properties of solids within the duration of the pulse,\cite{Schiffrin_at_al_Nature_2012_Current_in_Dielectric,Schultze_et_al_Nature_2012_Controlling_Dielectrics} which is a few femtosecond-long. 
The response of electron system of a solid to the optical field of the pulse strongly depends on the 
band structure of the solid.

For dielectrics, the main energy parameter, which determines the interaction of a solid with the laser pulse, is the bandgap 
$\Delta _g$ between the occupied valence band and the empty conduction band. 
If  the pulse frequency is small, $\omega _0 \ll \Delta _g /\hbar$, then the electron dynamics can be described in terms  of the dynamics of the passage through anticrossing points of quasistationary Wannier-Stark levels of conduction and valence bands in time dependent electric field of the laser pulse.\cite{Schiffrin_at_al_Nature_2012_Current_in_Dielectric,Schultze_et_al_Nature_2012_Controlling_Dielectrics,Dielectric_PRB_2012} The passage through such anticrossing points determines whether the electron dynamics is adiabatic or diabatic. The last anticrossing point corresponds to the electric field of the strength $F_{crit}$. Such field can  also be defined as the field which induces a change in electron
potential energy by  $\Delta_g$ over the lattice period $a\sim 5$ \AA. For silica with bandgap $\Delta \sim 10 $ eV, the critical field is $F_{crit} = \Delta _g/|e| a \sim 2 $ V/\AA. 
At such electric field, i.e. at the last anticrossing point of Wannier-Stark levels, the interband coupling is strong, which results in strong mixing of conduction and valence band states. Such mixing results in strong enhancement of dielectric response of the solid.\cite{Dielectric_PRB_2012} 

In addition to the enhancement of  dielectric susceptibility of the solid, the response of the electron 
system of dielectrics to a strong optical pulse shows another interesting property. 
Namely, the deviation of electron dynamics from adiabatic one results in finite charge 
transfer $\Delta Q$ through the system during the pulse.\cite{Schiffrin_at_al_Nature_2012_Current_in_Dielectric} For an ultrastrong pulse, the effective conductivity calculated from the transferred 
charge, $\sigma \sim \Delta Q/\tau _p F_0 $, is enhanced by almost 18 orders in magnitude compared to its   low-field value. The direction of the charge transfer is the same as the direction of the pulse maximum.   
For ultrashort laser pulse, the electron dynamics is also highly reversible, i.e. the electron system almost 
returns to its initial state after the pulse ends. Such reversibility was demonstrated both experimentally\cite{Schultze_et_al_Nature_2012_Controlling_Dielectrics} and numerically.\cite{Dielectric_PRB_2012} 
Thus, within the duration of ultrashort and strong laser pulse, the insulator shows strong enhancement of 
both dielectric response and electrical conductivity with highly reversible dynamics. 

In metals, where the conduction band is partially occupied, the main effect of interaction of 
ultrashort optical pulse with solid is strong modification of intraband electron dynamics.\cite{Metal_PRB_2013} The electron dynamics 
in strong optical pulse shows high frequency Bloch oscillations, which is visible in the generated electric current 
and in the shape of the optical pulse transmitted through the metal nanofilm.\cite{Metal_PRB_2013} In addition to such oscillations the 
highly nonlinear electron dynamics in ultrastrong optical pulse results in strong enhancement of the pulse
transmittance through the metal nanofilm.\cite{Metal_PRB_2013} Similar to dielectrics, the optical pulse also generates the transferred electric charge, but now the direction of the charge transfer is opposite to the direction of the 
pulse maximum. 

In the present paper we consider interaction of ultrashort  laser pulse with graphene monolayer\cite{Novoselov_nat_mater_2007, Electronic_properties_graphene_RMP_2009, graphene_advances_2010}. 
The purely two dimensional electron dynamics in graphene is characterized by unique dispersion 
relation,  the low energy part of which is relativistic with linear dependence of the electron energy on momentum. The behavior of such low energy electrons is described by the Dirac 
relativistic massless equation. Therefore, graphene is a semimetal with zero bandgap and relativistic low-energy dispersion. In this case the interaction of the laser pulse 
with graphene should show some similarity  to the behavior of a metal in strong optical pulse,
where the intraband electron dynamics determines the response of the electron system. 
Zero bandgap should also result in strong interband mixing of the states of the valence and conduction bands. 
Below we consider femtosecond-long laser pulses, for which the duration of the pulse is less 
than the electron scattering time, which is of the order of 1 ps.\cite{Graphene_scattering_time_PRB_2008} In this case the electron dynamics is coherent and is described by the time-dependent Schr\"odinger equation, where the time dependence is introduced 
through the time-dependent electric field of the optical pulse. 

The dynamics of graphene in long optical pulse with duration of hundred femtosecond, for which the scattering processes become important and the electron dynamics is incoherent, has been studied in Ref.\ \onlinecite{Graphene_THz_New_journal_2013} within the density matrix approach, where the sensitivity of the hot-electron Fermi distribution to the intensity of the optical pulse were reported. For long circular polarized optical pulses, the interaction of electrons in graphene with periodic electric field results also in formation the Floquet states and opening a gap in the energy spectrum of 
graphene\cite{Opening_gap_graphene_PRB_2009,Floquet_spectrum_graphene_PRL_2011,Tuning_gap_graphene_APL_2011} or graphene-like topological surface states of topological insulator.\cite{Floquet_TI_Science_2013}

\section{Model and Main Equations}

We consider an optical pulse, which is incident normally on graphene monolayer and has the following one-oscillation form     
\begin{equation}
F (t) = F_{0} e^{-u^2} \left( 1 - 2 u^2 \right),
\label{FV0}
\end{equation}
where $F_0$ is the amplitude, 
which is related to the pulse power ${\cal P} =c F_0^2/4 \pi$,  $c$ is speed of light,
$u = t/\tau $, and $\tau $ is the pulse length, which is set $\tau = 1 $ fs. 

\begin{figure}
\begin{center}\includegraphics[width=0.48\textwidth]{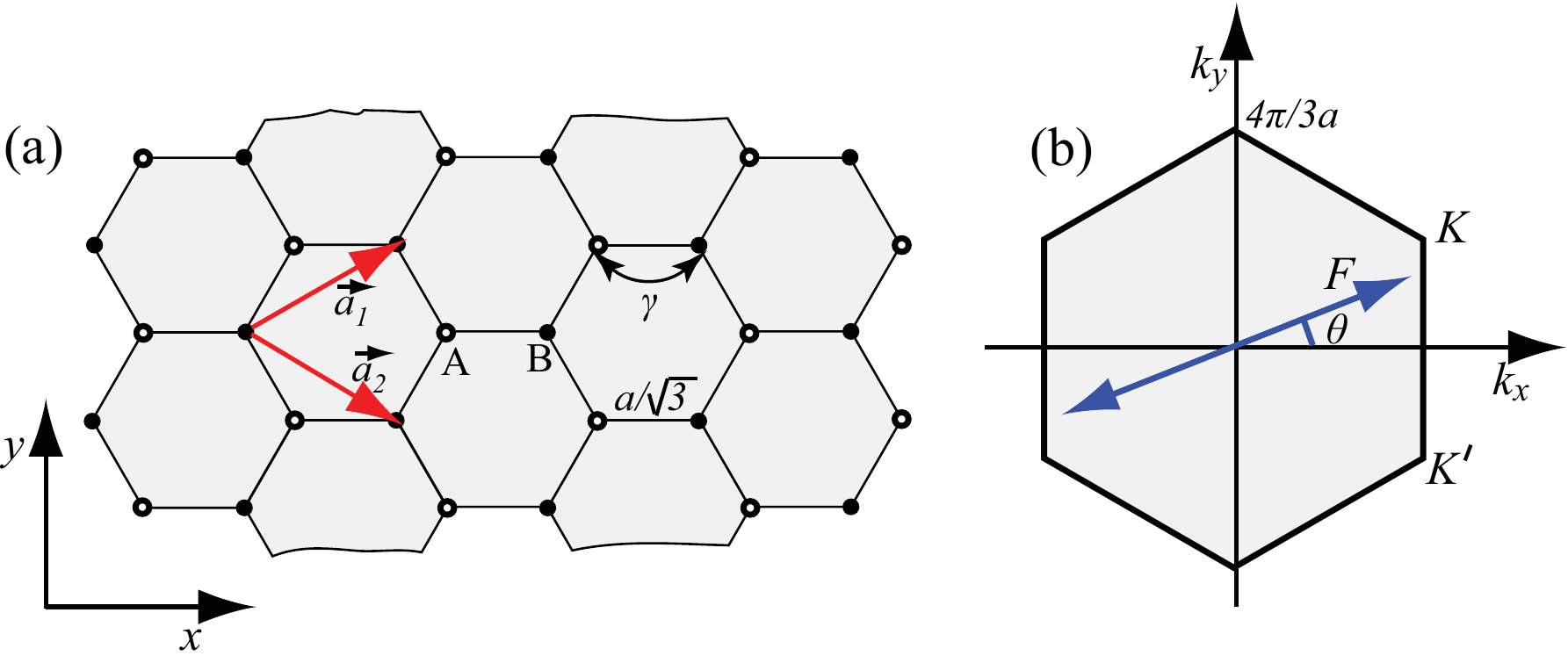}\end{center}
\caption{(a) Hexagonal lattice structure of 2D graphene. The graphene lattice consists of two 
inequivalent sublattices, which are labeled by "A" and "B". The vectors $\vec{a}_1=a/2(\sqrt{3},1)$ and $\vec{a}_2 = a/2(\sqrt{3},-1) $ are the direct lattice vectors of graphene. The nearest neighbor coupling, which is characterized by the hopping integral $\gamma $, is also shown. 
(b) The first Brillouin zone of reciprocal lattice of graphene. Points $K$ and $K^{\prime }$ are two degenerate Dirac points, corresponding to two valleys of low energy spectrum of graphene. Blue line with arrows shows polarization of the time-dependent electric field of the pulse. The polarization is characterized by angle $\theta $. 
} 
\label{graphene}
\end{figure}

We consider linearly polarized laser pulse, where the plane of polarization is characterized by 
angle $\theta $ measured relative to axis $x$. Here the $x$ and $y$ coordinate system is introduced in the plane of graphene and are determined by the crystallographic structure of graphene - see Fig.\ \ref{graphene}. The graphene has hexagonal lattice structure, which is 
shown in Fig.\ \ref{graphene}(a). The lattice has two sublattice, say "A" and "B", and is determined by two lattice vectors  $\vec{a}_1=a/2(\sqrt{3},1)$ and $\vec{a}_2 = a/2(\sqrt{3},-1) $, where 
$a = 2.46$ \AA \ is the lattice constant. The distance between the nearest neighbor atoms of graphene is $a/\sqrt{3}$. The first Brillouin zone of the reciprocal lattice of graphene, which is a hexagon, is shown in Fig.\ \ref{graphene}(b). The points $K= (2\pi/a) (1/3,1/\sqrt{3})$ and $K^{\prime}= (2\pi/a) (-1/3,1/\sqrt{3})$, which are the vertices of the hexagon, are the Dirac points. The energy gaps at these points are zero and the 
low energy spectra near these points are described by the Dirac relativistic equation. The points 
$K$ and $K^{\prime }$ correspond to two valley of low energy spectrum of graphene.

The Hamiltonian of an electron in graphene in the field of the optical pulse has the form 
\begin{equation}
{\cal H} = {\cal H}_0 + e \vec{F}(t) \vec{r} , 
\label{Htotal}
\end{equation}
where ${\cal H}_0$ is the field-free electron Hamiltonian, $\vec{r} =(x,y)$ is a 2D vector, and 
$\vec{F}(t) = [F(t)\cos\theta , F(t) \sin \theta]$. To describe both the conduction and 
valence bands of graphene within a single Hamiltonian, we consider the nearest neighbour 
tight-binding model of graphene. For such model the Hamiltonian ${\cal H}_0$ is the tight-binding Hamiltonian of graphene,\cite{Graphene_Wallace_PR_1947,Graphene_Weiss_PR_1958,graphene_Dresselhaus_1998,Carbon_nanotubes_2004} 
 which describes the tight-binding 
coupling between two sublattices "A" and "B" - see Fig.\ \ref{graphene}. In the reciprocal space the Hamiltonian 
${\cal H}_0$ is a $2\times 2$ matrix of the form\cite{Graphene_Wallace_PR_1947,Graphene_Weiss_PR_1958}
\begin{equation}
{\cal H}_0 =  
\left(
\begin{array}{cc}
0  &  \gamma  f(\vec{k}) \\
\gamma f^{*}(\vec{k})  & 0
\end{array}
\right)   ,
\end{equation}
where $\gamma=-3.03$ eV is the hopping integral and 
\begin{equation}
f(\vec{k}) = \exp \left( i \frac{a k_x} {\sqrt{3}}\right) + 2  \exp \left( -i \frac{a k_x} {2 \sqrt{3}}\right) 
 \cos \left( \frac{ak_y }{2} \right).
\end{equation}

The energy spectrum of Hamiltonian ${\cal H}_0$ consists of conduction band ($\pi^*$ or anti-bonding band) and valence bands ($\pi$ or bonding band) with the energy 
dispersion $E_{c}(\vec{k})= -\gamma |f(\vec{k})|$ (conduction band) and $E_{v}(\vec{k})= \gamma |f(\vec{k})|$ (valence band). The corresponding 
wave functions are 
\begin{equation}
\Psi^{(c)}_{\vec{k}} (\vec{r} ) = \frac{e^{i \vec{k} \vec{r}}}{\sqrt{2}}
\left( 
\begin{array}{c}
1 \\
e^{-i \phi_k }
\end{array}
\right)  
\label{functionV}
\end{equation}
and
\begin{equation}
\Psi^{(v)}_{\vec{k}} (\vec{r} ) = \frac{e^{i \vec{k} \vec{r}}}{\sqrt{2}}
\left( 
\begin{array}{c}
-1 \\
e^{-i \phi_k }
\end{array}
\right)  ,
\label{functionC}
\end{equation}
where $ f(\vec{k}) = |f(\vec{k}) | e^{i\phi _k}$. 
The wave functions $\Psi^{(c)}_{\vec{k}}$ and $\Psi^{(v)}_{\vec{k}}$  have two
components belonging to sublattices A and B, respectively.

When the duration of the laser pulse is less than the characteristic electron scattering time, which is 
around 1 ps \cite{Graphene_scattering_time_PRB_2008}, the electron dynamics in external electric field of the optical pulse is coherent and can be described 
by the time dependent Schr\"odinger equation
\begin{equation}
i\hbar \frac{d \Psi}{dt} = {\cal H} \Psi,
\label{time_eq1}
\end{equation} 
where the Hamiltonian (\ref{Htotal}) has explicit time dependence. 

The electric field of the optical pulse generates both interband and intraband electron dynamics. The 
interband dynamics introduces a coupling of the states of the conduction and valence bands and results in 
redistribution of electrons between two bands. For dielectrics, such dynamics results in its metallization, which 
manifest itself as a finite charge transfer through dielectrics and finite conduction band population after the 
pulse ends. 
 
It is convenient to describe the intraband dynamics, 
i.e. the electron dynamics within a single band, in the reciprocal space. In the reciprocal space, the 
electron dynamics is described by acceleration theorem, which has the following form
\begin{equation}
\hbar \frac{d\vec{k} }{dt} = e \vec{F}(t).  
\label{acceleration}
\end{equation}
The acceleration theorem is universal and does not depend on the dispersion law. Therefore the intraband electron dynamics is 
the same for both conduction and valence bands. For an electron 
with initial momentum $\vec{q}$ the electron dynamics is 
described by the time dependent wave vector, $\vec{k}_T(\vec{q},t)$, which is given by the solution of Eq.\ (\ref{acceleration}),
\begin{equation} 
\vec{k}_T(\vec{q},t) = \vec{q} +  
\frac{e}{\hbar} \int^t_{-\infty}  \vec{F} (t_1) dt_1 .
\label{kT}
\end{equation}
The corresponding wave functions are the Houston functions, 
\cite{Houston_PR_1940_Electron_Acceleration_in_Lattice} $\Phi^{(H)}_{\alpha q }(\vec{r}, t)$,
\begin{equation}
\Phi^{(H)}_{\alpha \vec{q} }(\vec{r},t) = \Psi^{(\alpha)}_{\vec{k}_T(\vec{q},t)} (\vec{r} )
e^{- \frac{i}{\hbar }   \int ^t_{-\infty} \!\! dt_1 E_{ \alpha} [\vec{k}_T(\vec{q},t_1) ] }, 
\label{phi_h}
\end{equation}
where $\alpha = v$ (valence band) or $\alpha = c$ (conduction band). 

Using the Houston functions as the basis, we express the general solution of the time-dependent
Schr\"odinger equation (\ref{time_eq1}) in the following form
\begin{equation}
\Psi_{\vec{q}} (\vec{r}, t) = \sum_{\alpha = v,c} \beta _{\alpha \vec{q} } (t) 
                  \Phi^{(H)}_{\alpha \vec{q} }(\vec{r},t).
 \label{psi0}
\end{equation}
The solution (\ref{psi0}) is parametrized by initial electron wave vector $\vec{q}$. Due to 
universal electron dynamics in the reciprocal space, the states, which belong to different bands (conduction 
and valence bands) and which have the same initial wave vector $\vec{q}$, will have the same wave vector 
$\vec{k}_T(\vec{q},t) $ at later moment of time $t$. Since the interband dipole matrix element, which 
determines the coupling of the  conduction and valence band states in external electric field, 
is diagonal in the reciprocal space, then the states with different initial wave vectors are not coupled by 
the pulse field. As a result in Eq.\ (\ref{psi0}), for each value of initial wave vector 
$q$, we need to find only two time-dependent expansion coefficients $\beta _{v \vec{q} } (t)$  and $\beta _{c \vec{q} } (t)$.
Such decoupling of the states with different values of $\vec{q}$ is the property of coherent 
dynamics. For incoherent dynamics, the electron scattering couples the states with different wave vectors $\vec{q}$. In this case the dynamics is described by the density matrix. 

The expansion coefficients satisfy the following system of differential equations 
\begin{eqnarray}
& & 
\frac{d \beta _{c \vec{q} } (t)}{dt} = -i \frac{\vec{F}(t) \vec{Q} _{\vec{q}} (t) }{\hbar } 
  \beta _{v \vec{q} } (t),    \label{system1} \\
& & 
\frac{d \beta _{v \vec{q} } (t)}{dt} = -i \frac{\vec{F}(t)  \vec{Q}^* _{\vec{q}} (t) }{\hbar } 
  \beta _{c \vec{q} } (t),
\label{system2}
\end{eqnarray}
where the vector-function $\vec{Q} _{\vec{q}} (t)$ is proportional to the interband dipole matrix element 
\begin{equation}
\vec{Q} _{\vec{q}} (t) =  \vec{D}[\vec{k}_T(\vec{q},t)]   
   e^{- \frac{i}{\hbar }   \int ^t_{-\infty} \!\! dt_1 
       \left\{  E_{c} [\vec{k}_T(\vec{q},t_1)]  -E_{v} [\vec{k}_T(\vec{q},t_1)]  \right\}  },
\label{Q_vector}
\end{equation}
where $\vec{D}(\vec{k} ) =[D_x(\vec{k}),D_y(\vec{k})]  $ is the dipole matrix element between the states of the conduction and 
valence bands with wave vector $\vec{k}$, i.e. 
\begin{equation}
\vec{D}(\vec{k} ) = \left\langle \Psi^{(c)}_{\vec{k} }\right|  e\vec{r} 
\left| \Psi^{(v)}_{\vec{k} }\right\rangle.
\label{dipole}
\end{equation}
Substituting the conduction and valence band wave functions (\ref{functionV}) and 
(\ref{functionC}) into Eq.\ (\ref{dipole}), we obtain the following 
expressions for the interband dipole matrix elements 
\begin{equation}
D_x(\vec{k})  = \frac{e a}{2\sqrt{3}}  \frac{1+ \cos \left( \frac{ak_y}{2}  \right) 
       \left[ \cos \left(\frac{3ak_x}{2\sqrt{3} } \right) - 2\cos \left(\frac{ak_y}{2}  \right)   \right]  }
         {1+4 \cos \left( \frac{ak_y}{2}  \right) 
       \left[ \cos \left(\frac{3ak_x}{2\sqrt{3} } \right) + \cos \left(\frac{ak_y}{2}  \right)   \right]  } 
\label{Dx}
\end{equation}
and 
\begin{equation}
D_y(\vec{k})  = \frac{e a}{2}  \frac{\sin \left( \frac{ak_y}{2}  \right)  
      \sin \left(\frac{3ak_x}{2\sqrt{3} } \right) }
         {1+4 \cos \left( \frac{ak_y}{2}  \right) 
       \left[ \cos \left(\frac{3ak_x}{2\sqrt{3} } \right) + \cos \left(\frac{ak_y}{2}  \right)   \right]  }. 
\label{Dy}
\end{equation}
The system of equations (\ref{system1})-(\ref{system2}) describes the interband electron dynamics and determines the mixing of the 
conduction band and the valence band states in the electric field of the pulse.  
There are two solutions of the system (\ref{system1})-(\ref{system2}), which correspond to two initial conditions: $(\beta _{v \vec{q} },
\beta _{c \vec{q} } )  = (1,0)$ and  $(\beta _{v \vec{q} },
\beta _{c \vec{q} } ) = (0,1)$. These solutions 
determine the evolution of the states, which are initially in the valence band or in the conduction band, respectively.

For undoped graphene all states of the valence band are occupied 
and all states of the conduction band are empty. For an electron, which is initially in the valence band the mixing of the states of different bands is characterized by the time-dependent component  $|\beta _{c \vec{q} }(t)|^2$. We can also define the time-dependent 
total occupation of the conduction band for undoped graphene from the 
following expression 
\begin{equation}
{\cal N}_c (t) = \sum_{\vec{q}} |\beta _{c \vec{q} }(t)|^2,
\label{Ntotal}
\end{equation}
where the sum is over the first Bruilluen zone and 
the solution $\beta _{c \vec{q} }(t)$ in Eq.\ (\ref{Ntotal}) satisfies 
the initial condition  $(\beta _{v \vec{q} }, \beta _{c \vec{q} } )  = (1,0)$. 

Redistribution of electrons between the conduction and the valence bands in time-dependent electric field also generates electric current, 
which can be calculated in terms of the operator of velocity 
from the following expression 
\begin{equation}
J_j (t) = \frac{e}{a^2} \sum_{\vec{q}} \sum_{\alpha_1 =v,c} 
\sum_{\alpha _2 = v,c}
\beta _{\alpha_1 \vec{q}}^* (t) 
    {\cal V}_{j}^{\alpha_1 \alpha_2}  \beta _{\alpha_2\vec{q}} (t),
\label{current}
\end{equation}
where $j = x,y$ and $ {\cal V}_{j}^{\alpha_1 \alpha_2} $ are 
the matrix elements of the velocity operator 
$\hat {\cal V}_{j} = \frac{1}{\hbar } 
\frac{\partial {\cal H}_0}{ \partial k_j}$ between the conduction 
and valence band states. With the known wave functions (\ref{functionV})-(\ref{functionC}) of the 
conduction and valence bands the matrix elements of the velocity 
operator are 
\begin{eqnarray}
 {\cal V}_x^{cc} = -{\cal V}_x^{vv} = \frac{a\gamma }{\sqrt{3} \hbar} & &
\left[  \sin\left( \frac{ak_x}{\sqrt{3}} - \phi_{\vec{k}}   \right)
\right.  +  \nonumber \\
& & \left.   \sin\left( \frac{ak_x}{\sqrt{3}} + \phi_{\vec{k}}   \right)  \cos \frac{ak_y}{2} \right],
\end{eqnarray}
\begin{equation}
{\cal V}_y^{cc} = -{\cal V}_y^{vv} = \frac{a\gamma}{\hbar}    \cos\left( \frac{ak_x}{2\sqrt{3}} + \phi_{\vec{k}}   \right)   \sin \frac{ak_y}{2} ,
\end{equation}
\begin{eqnarray}
{\cal V}_x^{cv} =  -i \frac{2a\gamma}{\sqrt{3} \hbar} & & 
\left[   \cos\left( \frac{ak_x}{\sqrt{3}} - \phi_{\vec{k}}   \right)  -
\right.   \nonumber \\
& &  \left. \cos\left( \frac{ak_x}{\sqrt{3}} + \phi_{\vec{k}}   \right)  \cos \frac{ak_y}{2} \right],
\end{eqnarray}
and
\begin{equation}
{\cal V}_y^{cv} =  -i \frac{2a\gamma}{\hbar}  \sin\left( \frac{ak_x}{\sqrt{3}} + \phi_{\vec{k}}   \right)  \cos \frac{ak_y}{2}. 
\end{equation}
The interband matrix elements of the velocity operator, ${\cal V}_x^{cv}$ and 
${\cal V}_y^{cv}$, are related to the interband dipole matrix elements, 
${\cal V}_x^{cv} = i D_x(\vec{k})  \left[  E_c (\vec{k}) - E_v (\vec{k})  \right]/\hbar $ and ${\cal V}_y^{cv} = i D_y(\vec{k})  \left[  E_c (\vec{k}) - E_v (\vec{k})  \right]/\hbar $. \cite{Landau_Lifshitz_Quantum_Mechanics:1965}

Within the nearest neighbor tight binding model, the graphene has electron-hole symmetry, which results in the relation ${\cal V}_y^{cc} = -{\cal V}_y^{vv}$.
Inclusion into the model the higher order tight-binding couplings, e.g. next-nearest neighbor terms, introduced electron-hole asymmetry, which results in different magnitudes of velocity in the conduction and valence bands.\cite{Tight_binding_MacDonald_PRB_2013} This asymmetry is weak and does not 
change the main results presented below. 

If the direction of electric field of the pulse is  
along the direction of high symmetry of graphene crystal, then 
the current (\ref{current}) is generated along the direction of electric field of the pulse only, $J_{||}$. For graphene, the directions of high 
symmetry correspond to polarization angles $\theta =0$ and $30^0$. 
If polarization of electric field is not along the direction of 
high symmetry of graphene, then the current is generated in both the 
direction of the field, $J_{||}$, and in the direction perpendicular to the field, $J_{\perp }$. 

The generated current results in charge transfer through the system, which is determined by an expression 
\begin{equation}
Q_{tr,\mu } = \int_{-\infty}^{\infty }  dt J_{\mu } (t),
\label{Qtr}
\end{equation}
where $\mu = ||$ or $\perp $, which corresponds to the charge transfer along the direction of polarization of the 
laser pulse and in the direction perpendicular to polarization of the pulse, respectively. 
The transferred charge is nonzero only due to irreversibility of electron dynamics in the optical pulse. For completely reversible dynamics, when the system returns to its initial state, the transferred charge is exactly zero. Indeed, since the current can be expressed in terms of polarization $\vec{P}(t)$ of the electron system as $\vec{J} (t)= d \vec{P}(t)/dt$, then the transferred charge is determined by the residual polarization of the system, i.e. polarization of the electron system after the pulse ends, $Q_{tr,\mu } = P_{\mu }(t\rightarrow \infty )$. The residual population is nonzero only for irreversible dynamics.

\section{Results and Discussion}

\subsection{Interband coupling}
\label{interband}

\begin{figure}
\begin{center}\includegraphics[width=0.46\textwidth]{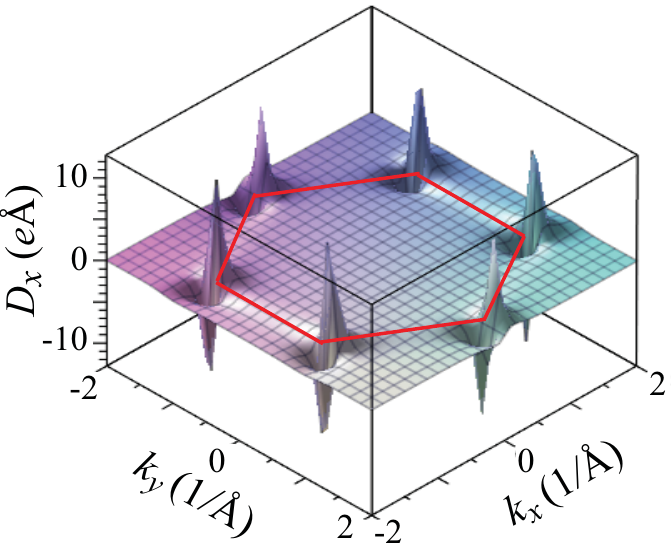}\end{center}
\caption{Interband dipole matrix element $D_x$ is shown as a function 
of the wave vector $\vec{k}$. 
The red lines show the boundary of the first Brillouin zone. The 
dipole matrix element is singular near the Dirac points ($K$ and $K^{\prime}$ points).    
} 
\label{z_x_y}
\end{figure}

The electron dynamics in time dependent electric field is determined by two unique properties of graphene: (i) zero band gap, which results in 
strong interband mixing even in a weak electric field, and (ii) strong dependence of interband dipole matrix elements on the wave vector. 
The interband dipole matrix elements, $D_x$ and $D_y$, are singular at the Dirac points, $K$ and $K^{\prime }$. Near these points the dipole matrix elements behave as $\sim 1/\Delta k$, where $\Delta k = |k - k_K|$ is the deviation of the wave vector from its value at the 
nearest Dirac point. The  dipole matrix element $D_x$ calculated from Eq.\ \ref{Dx} is shown as a function of the wave vector  in Fig. \ref{z_x_y}. The dipole matrix element become large near the Dirac points. The dipole matrix element $D_y$ has similar behavior. 

Away from the Dirac points the dipole matrix elements have a typical value of $ea/2\approx 1.2 e$\AA. At the center of the Bruilluen zone, 
i.e. at $\vec{k} = 0$, the dipole matrix elements are zero, i.e. at this point there is no interband coupling. 

In an electric field, either constant or time-dependent, the 
electron dynamics within a single band can be described in terms of 
time-dependent wave vector $\vec{k}_T(\vec{q},t)$, which introduces 
an electron trajectory in the reciprocal space. Then the effective interband coupling is determined by the average value of dipole matrix elements along the electron trajectory.

\subsection{Conduction band population}

One of the characteristics of electron dynamics in time-dependent electric field of the
optical pulse is redistribution of  electrons between the conduction and valence 
band states. In undoped graphene, the valence band is initially fully occupied, while the 
conduction band is empty. The electric field introduces coupling of the states of conduction and 
valence bands, which results in finite population of the conduction band. 

\begin{figure}
\begin{center}\includegraphics[width=0.4\textwidth,angle =90]{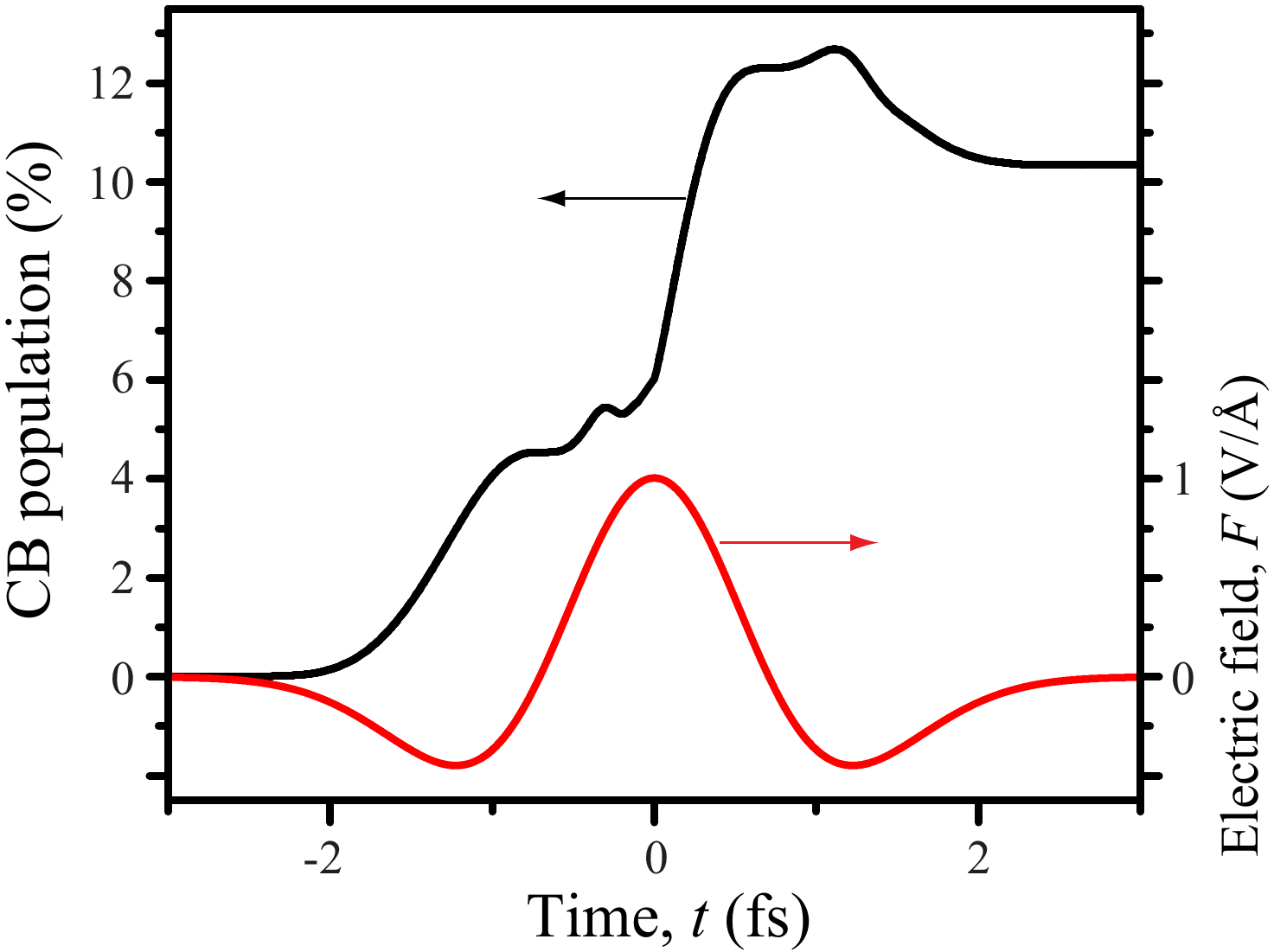}\end{center}
\caption{Time-dependent conduction band population, ${\cal N}_c(t)$, and the corresponding 
time-dependent electric field of the laser pulse are shown. The polarization of the pulse is 
along axis $x$, i.e. $\theta = 0$. 
} 
\label{CB_field}
\end{figure}

At first, we characterize the redistribution of the electrons between the valence and conduction bands in terms of the total population of the conduction band, ${\cal N}_c (t)$, - see Eq.\ (\ref{Ntotal}). The time dependence of the conduction band population ${\cal N}_c(t)$ is correlated with the time-dependent electric field.  
In Fig.\ \ref{CB_field} the conduction band population ${\cal N}_c(t)$ is shown as a function of time 
together with corresponding time-dependent electric field. Specific feature of this dependence is $\approx \pi/2$
phase shift between the conduction band population ${\cal N}_c(t)$ and the electric field $F(t)$. 
The maxima  of the conduction band population correspond to zeros of the electric field. Such phase shift 
between ${\cal N}_c(t)$ and $F(t)$ is due to strong wave vector dependence of the interband coupling. For insulators, where the interband coupling has weak dependence 
on the wave vector, the maxima of the conduction band population correspond to the maxima of the absolute value of electric field 
$|F(t)|$.\cite{Dielectric_PRB_2012}

The time-dependence of the conduction band population also illustrates the fact that the electron dynamics is 
highly irreversible, i.e. the electron system does not return to its original state after the pulse ends. 
The residual conduction band population, i.e. population after the pulse ends, is large and comparable to 
the maximum conduction band population during the pulse. 

\begin{figure}
\begin{center}\includegraphics[width=0.4\textwidth]{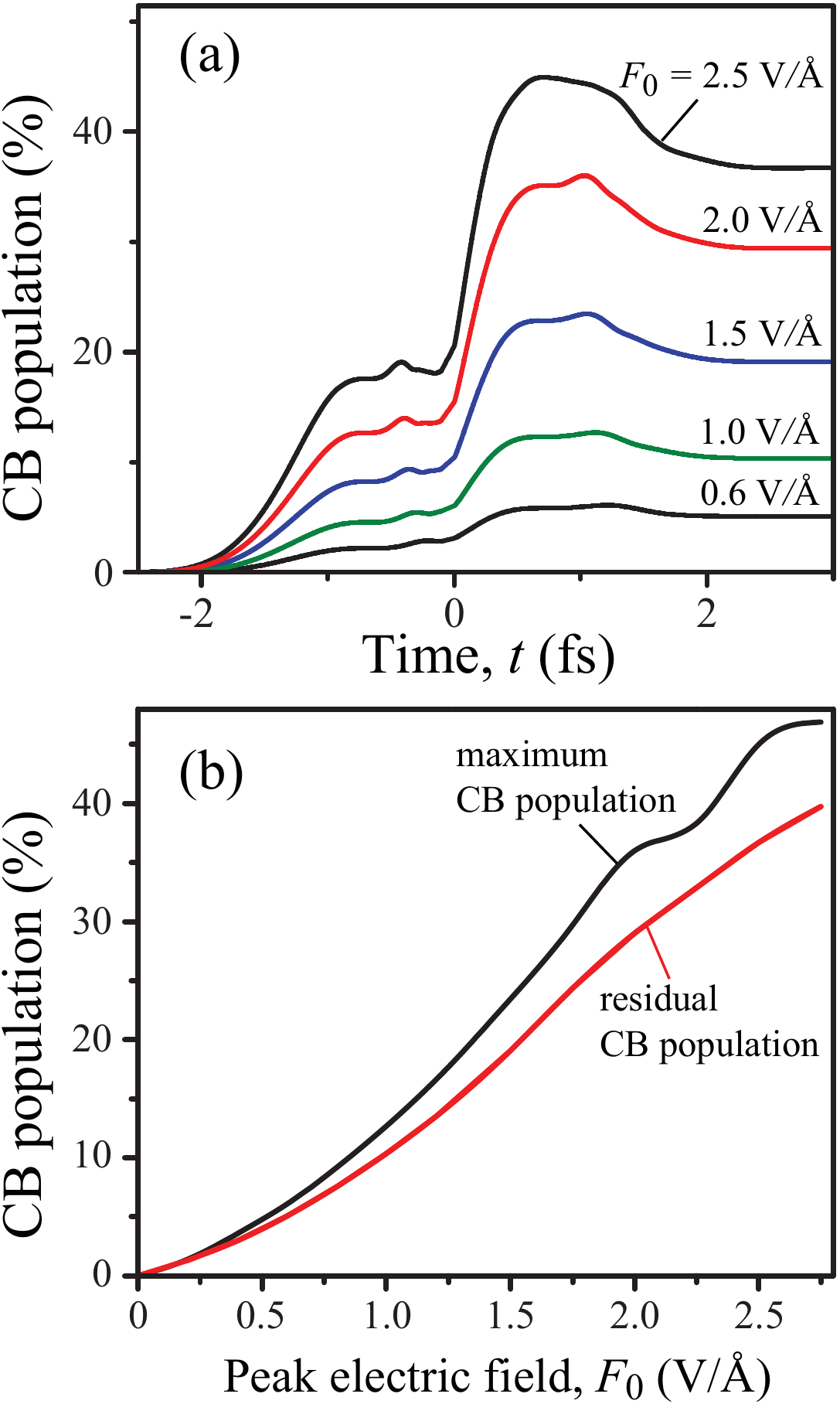}\end{center}
\caption{(a) Time-dependent conduction band population, ${\cal N}_c(t)$, is shown for different 
intensities (peak electric fields $F_0$) of the optical pulse. (b) The maximum and the residual conduction band populations
are shown as a function of the pulse amplitude $F_0$. The polarization of the pulse is 
along axis $x$, i.e. $\theta = 0$.    
} 
\label{CBs_maximum}
\end{figure}

Irreversible dynamics of electron system  persists at all pulse intensities. This property is illustrated in 
Fig.\ \ref{CBs_maximum}, where the conduction band population is shown for different pulse amplitudes, $F_0$. 
The time-dependent conduction band population, ${\cal N}_c(t)$, have the similar 
time profile for all pulse amplitudes. The maximum conduction band population is realized at $t\approx 1$ fs, 
which corresponds to the last local maximum of the magnitude of electric field, $|F(t)|$. At all pulse 
amplitudes, the residual conduction band population, ${\cal N}(t\rightarrow \infty)$, is comparable to the 
maximum population. Both the residual and the maximum populations monotonically increase with increasing 
peak electric field $F_0$, see Fig.\  \ref{CBs_maximum}(b). The results shown in Figs.\ \ref{CB_field} and 
\ref{CBs_maximum} correspond to polarization of the optical pulse along the $x$ axis, i.e. $\theta = 0$. The conduction band 
population has weak dependence on the polarization of the optical pulse, i.e. on the value of angle $\theta$, 
and the results similar to Figs.\ \ref{CB_field} and \ref{CBs_maximum} are valid for other angles $\theta $.

\begin{figure}
\begin{center}\includegraphics[width=0.5\textwidth]{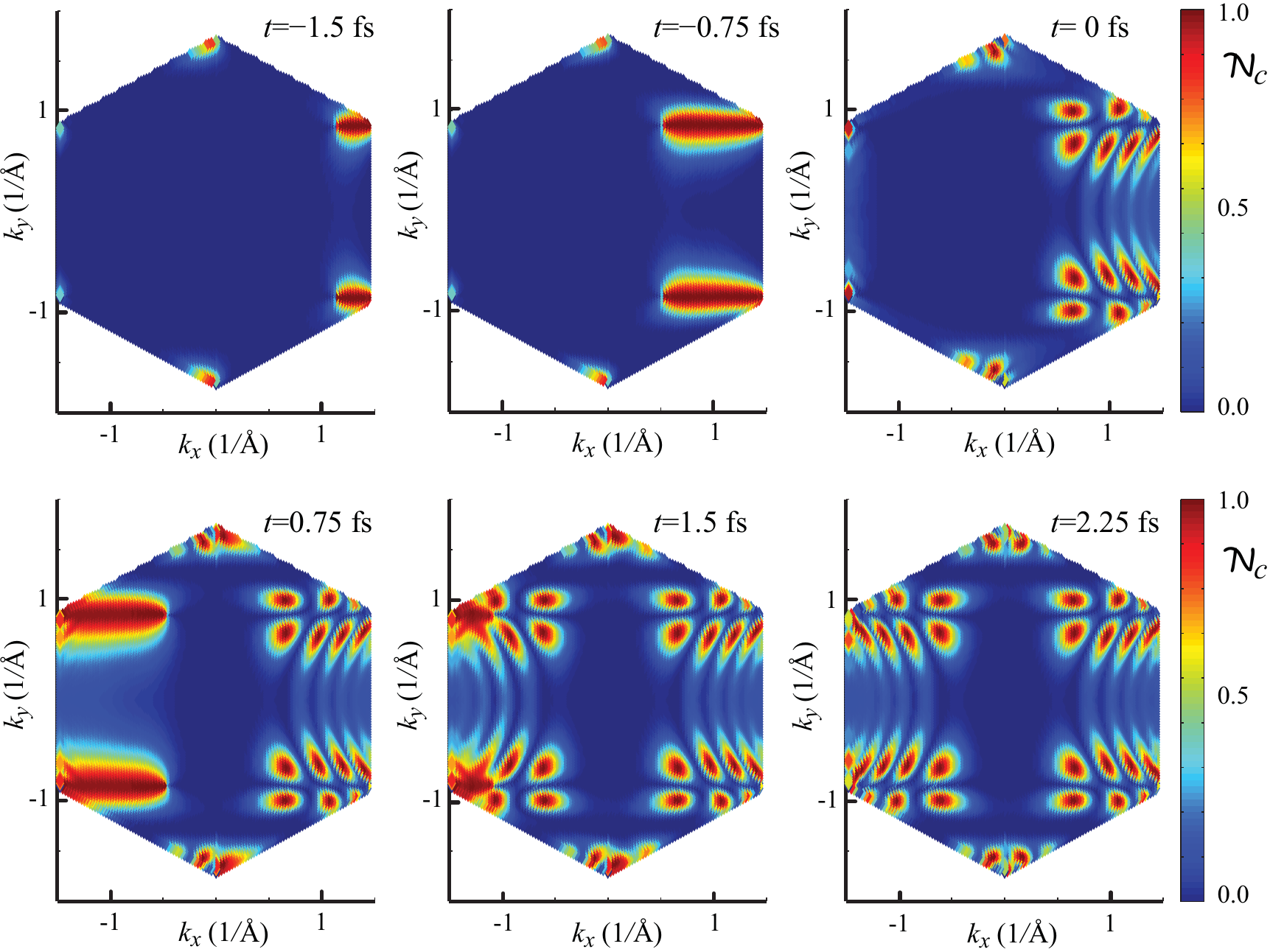}\end{center}
\caption{The conduction band population $|\beta_{c\vec{k}}(t)|^2$ is shown as a function of the wave vector at 
different moments of time. Only the first Bruilluen zone of the 
reciprocal space is shown. The peak electric field of the pulse 
is $F_0 = 1$ V/\AA. Different colors correspond to different values of the conduction band population as shown in 
the figure. 
} 
\label{Population_KxKy_time}
\end{figure}

The irreversible electron dynamics and the phase shift between the time-dependent conduction band 
population and the time-dependent electric field are due to gapless energy dispersion in graphene and 
strong dependence of the interband dipole matrix elements, $D_x$ and $D_y$, on the wave vector. Singularity of the dipole matrix elements at the Dirac points also results in strongly nonuniform distribution of the 
conduction band population in the reciprocal space. Such distribution is given by the function $|\beta_{c\vec{q}}(t)|^2$ and has strong dependence on the wave vector $\vec{q}$. In Fig.\ \ref{Population_KxKy_time} the 
conduction band population is shown in the reciprocal space at different moments of time. 
 There are strongly localized regions in the reciprocal space with 
the conduction band population equals almost 1. Therefore at these points the electrons are completely transferred 
to the conduction band with zero population of the valence band. The regions with high conduction band population 
evolve with time. Such evolution correlates with time dynamics of electrons in the reciprocal space, which is determined by the time-dependent wave vector $\vec{k}_T$. The regions of high conduction band population are 
concentrated near the Dirac points and their arrangement clearly follows the polarization of the optical 
pulse, which in Fig.\ \ref{Population_KxKy_time} is along axis $x$. The dynamics of 
formation of localized regions of high conduction band population is also shown in Fig.\ \ref{Population_KxKy_time}. Initially the conduction band population is high within large region of an oval shape
in the reciprocal space  (see distribution at $t=-0.75$ fs). Then this oval shape becomes broken into small 
localized regions with high values of the conduction band population.

\begin{figure}
\begin{center}\includegraphics[width=0.4\textwidth]{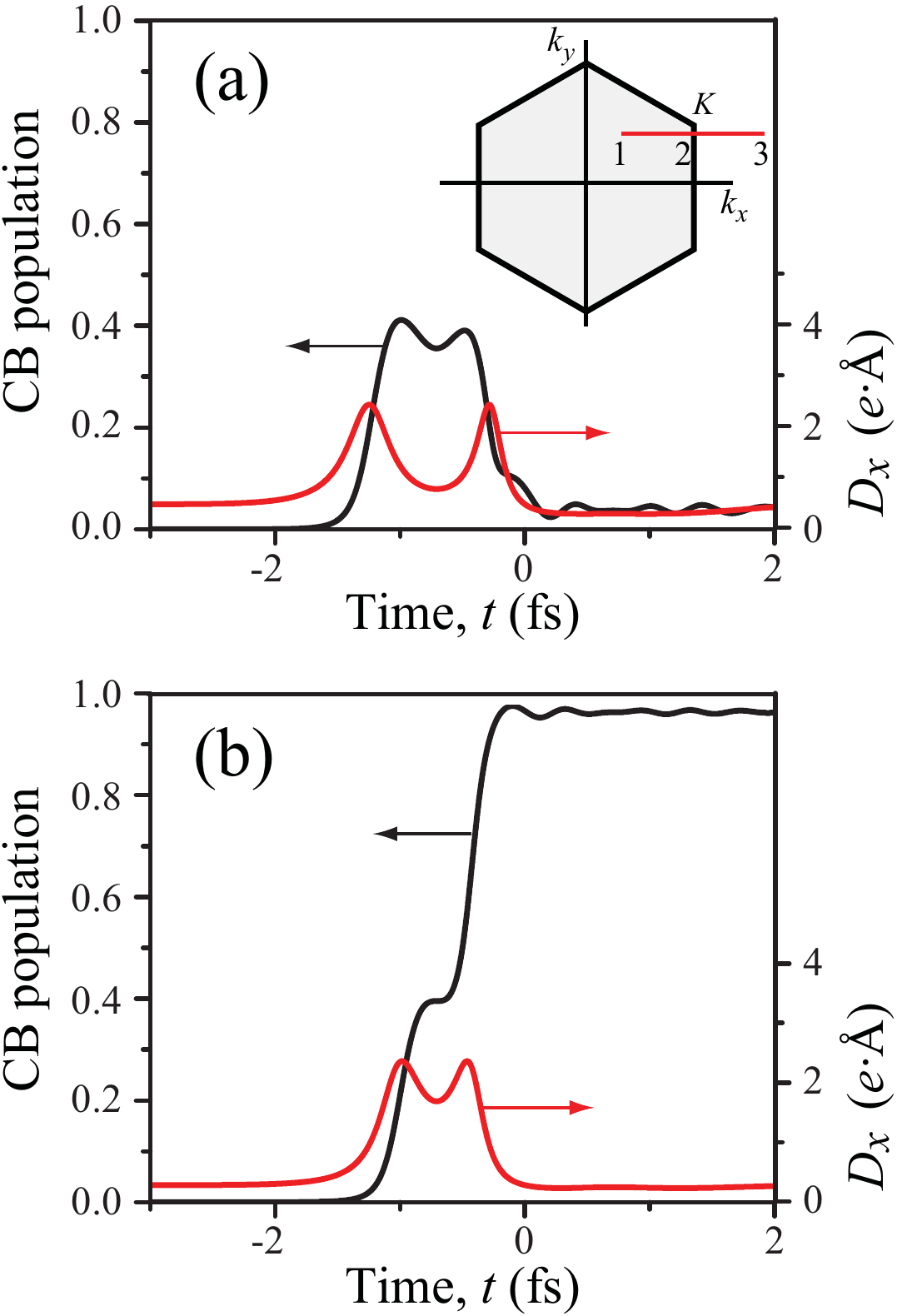}\end{center}
\caption{
Time-dependent conduction band population and the corresponding dipole 
matrix element $D_x$ are shown for some initial wave vector $\vec{q}$ of the reciprocal space. The conduction band population is calculated as $|\beta _{c\vec{q}}(t)|^2$ and the dipole matrix element is defined as $D_x (\vec{k}_T(\vec{q},t)$. Two different initial wave vectors 
in panels (a) and (b) correspond to small and large residual conduction band populations, respectively. The inset in panel (a) illustrates 
schematically the electron dynamics in the reciprocal space: the electron is transferred along the path "1"$\rightarrow$"2"$\rightarrow$"3"$\rightarrow$"2"$\rightarrow$"1". The polarization of the optical pulse is along axis $x$.  
} 
\label{CB_single_q}
\end{figure}

The formation of the localized regions with high conduction band population is due to singularity of the intraband dipole matrix elements at the Dirac points.  The 
interband dipole matrix elements are large near the Dirac points and are diverging exactly at the Dirac points - see section \ref{interband}.
An electron with initial wave vector $\vec{q}$ propagates in the reciprocal space along the direction of the electric field and the electron wave vector at moment of time $t$ is given by the function  
$\vec{k}_T (\vec{q},t)$, see Eq.\ (\ref{kT}). The trajectory of such electron is shown schematically in the 
inset in Fig.\ \ref{CB_single_q}(a), where the electron, which is initially at point "1", is transferred along 
the path "1"$\rightarrow$"2"$\rightarrow$"3"$\rightarrow$"2"$\rightarrow$"1" during the pulse. Since the area under the pulse is zero, the electron returns to the initial point "1".  
Along this closed path the interband coupling, which is proportional to the interband dipole matrix element, is the 
strongest near the point closest to the Dirac points, i.e. near point "2". Thus, the strongest mixing of the 
states of the conduction and valence bands occurs when the electron passes through point "2". For the closed 
path "1"$\rightarrow$"3"$\rightarrow$"1" there are two passages of point "2". As a result there are two strong 
changes in the conduction band population. These two changes can be constructive or destructive, resulting 
in final large or small conduction band population, respectively. These two possibilities are shown in 
Fig.\ \ref{CB_single_q}, where the time-dependent conduction band population is shown for two initial wave vectors
$\vec{q}$. The time-dependent interband dipole matrix element, $D_x$, calculated at wave vector 
$\vec{k}_{T}(\vec{q},t)$ is also shown in Fig.\ \ref{CB_single_q}. The two maxima in the time-dependent 
dipole matrix element correspond to two passages of the point "2" shown in the inset in Fig.\ \ref{CB_single_q}(a). 
For both initial wave vectors [see Fig.\ \ref{CB_single_q} (a) and (b)] the maxima of the dipole matrix element are correlated with large changes in the conduction band population. In Fig.\ \ref{CB_single_q}(b) these changes are constructive resulting in large conduction band population after the pulse ends, while in Fig.\ \ref{CB_single_q}(a) the changes are destructive, which results in small final conduction band population. 
Whether changes of the conduction band population constructive or destructive is determined by the phase 
accumulated between two consecutive passages of point "2". The phase is determined by exponential factor 
in the expression (\ref{Q_vector}) for the vector-function $\vec{Q} _{\vec{q}} (t)$.

\begin{figure}
\begin{center}\includegraphics[width=0.4\textwidth]{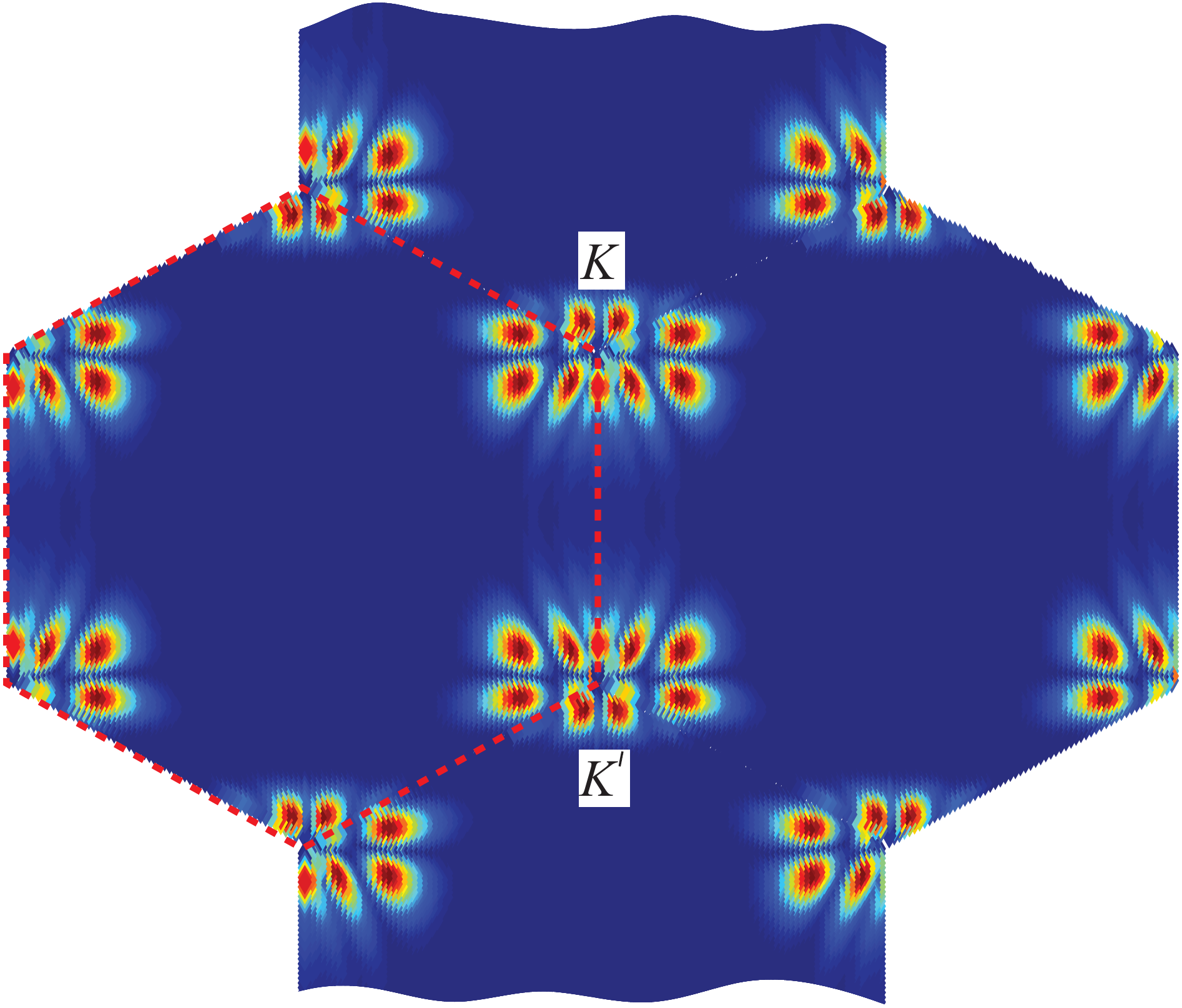}\end{center}
\caption{Residual conduction band population (population after the pulse ends) is shown as a function of the wave vector near the Dirac points: points $K$ and $K^{\prime }$, which correspond to two valley 
of graphene. The red dotted line shows the boundary of the first Brillouin zone. Different colors correspond to different values of the conduction band population as shown in the figure. Polarization of the optical pulse is along axis $x$. 
} 
\label{PopulationNearKpoints}
\end{figure}

\begin{figure}
\begin{center}\includegraphics[width=0.5\textwidth]{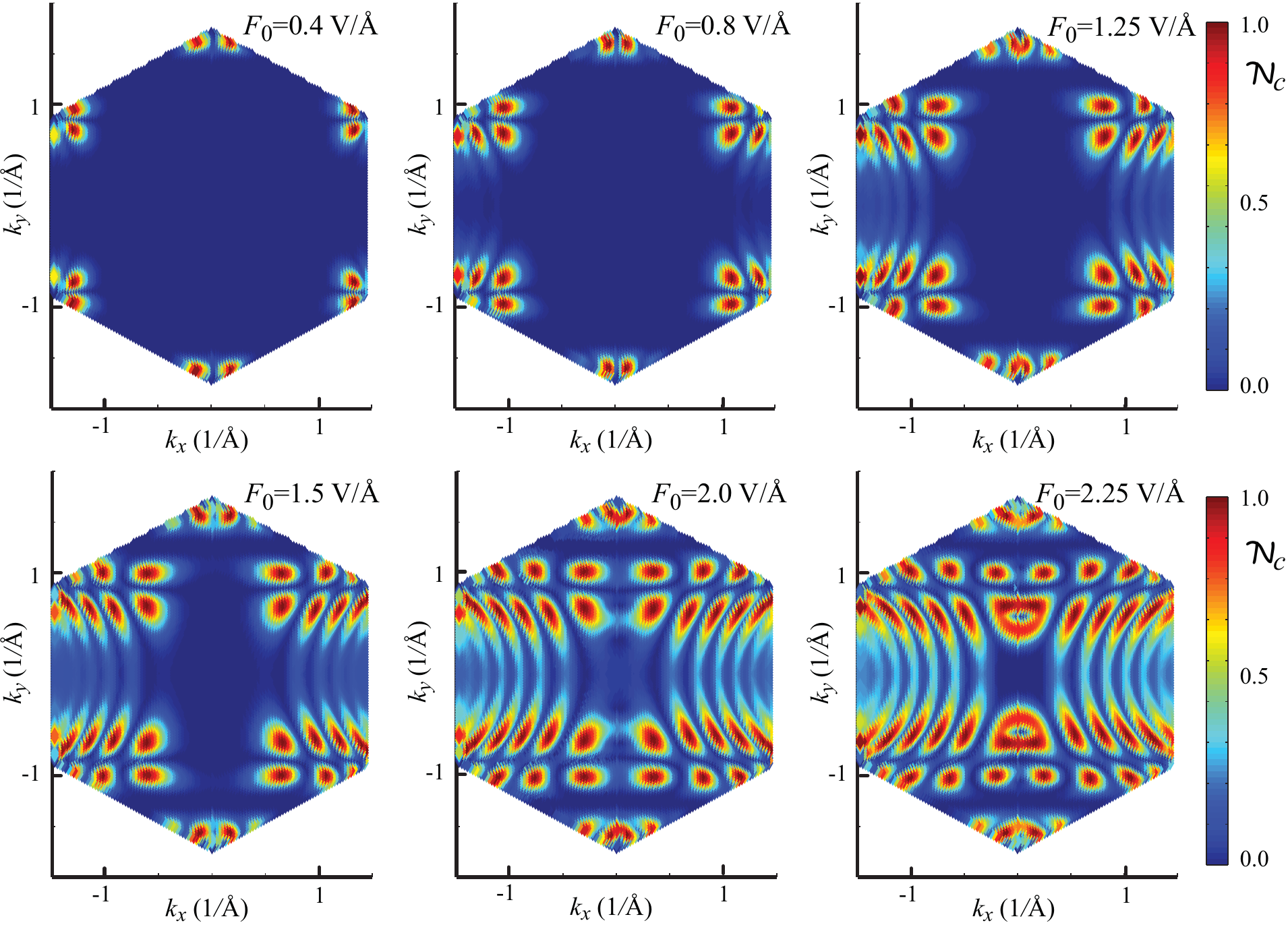}\end{center}
\caption{Residual conduction band population $|\beta_{c\vec{k}}(t\rightarrow \infty )|^2$ is shown as a function of the wave vector at 
different amplitudes $F_0$ of the optical pulse. Only the first Bruilluen zone of the 
reciprocal space is shown. The polarization of electric field is along axis $x$. Different colors correspond to different values of the conduction band population as shown in 
the figure.     
} 
\label{Population_KxKy_field}
\end{figure}

After the pulse ends the localized spots of high conduction band population are accumulated near the 
Dirac $K$ and $K^{\prime}$ points, which is illustrated in Fig.\ \ref{PopulationNearKpoints}, where the 
results shown in Fig.\ \ref{Population_KxKy_time} are redrawn beyond the first Brillouin zone.
The spots of high  conduction band population form two parallel arrays oriented along axis $x$, i.e. along 
the direction of the electric field. The number of spots in each array depends on the intensity of the optical pulse. 
In Fig.\ \ref{Population_KxKy_field} the residual conduction band population is shown as a function of wave vector 
$\vec{k}$ for different amplitudes $F_0$ of the optical pulse. In all cases the structure of two arrays of 
high conduction population spots near the Dirac points persists but the number of spots increases with $F_0$. 
This is due to the fact that with increasing the pulse amplitude the electron is transferred over a longer 
distance in the reciprocal space and the farther points in the reciprocal space can reach the Dirac points, where 
the strong interband coupling is realized. In terms of the schematic diagram shown in the inset in Fig.\ \ref{CB_single_q}(a) it means that the distance between points "1" and "2" increases with pulse amplitude.

\begin{figure}
\begin{center}\includegraphics[width=0.48\textwidth]{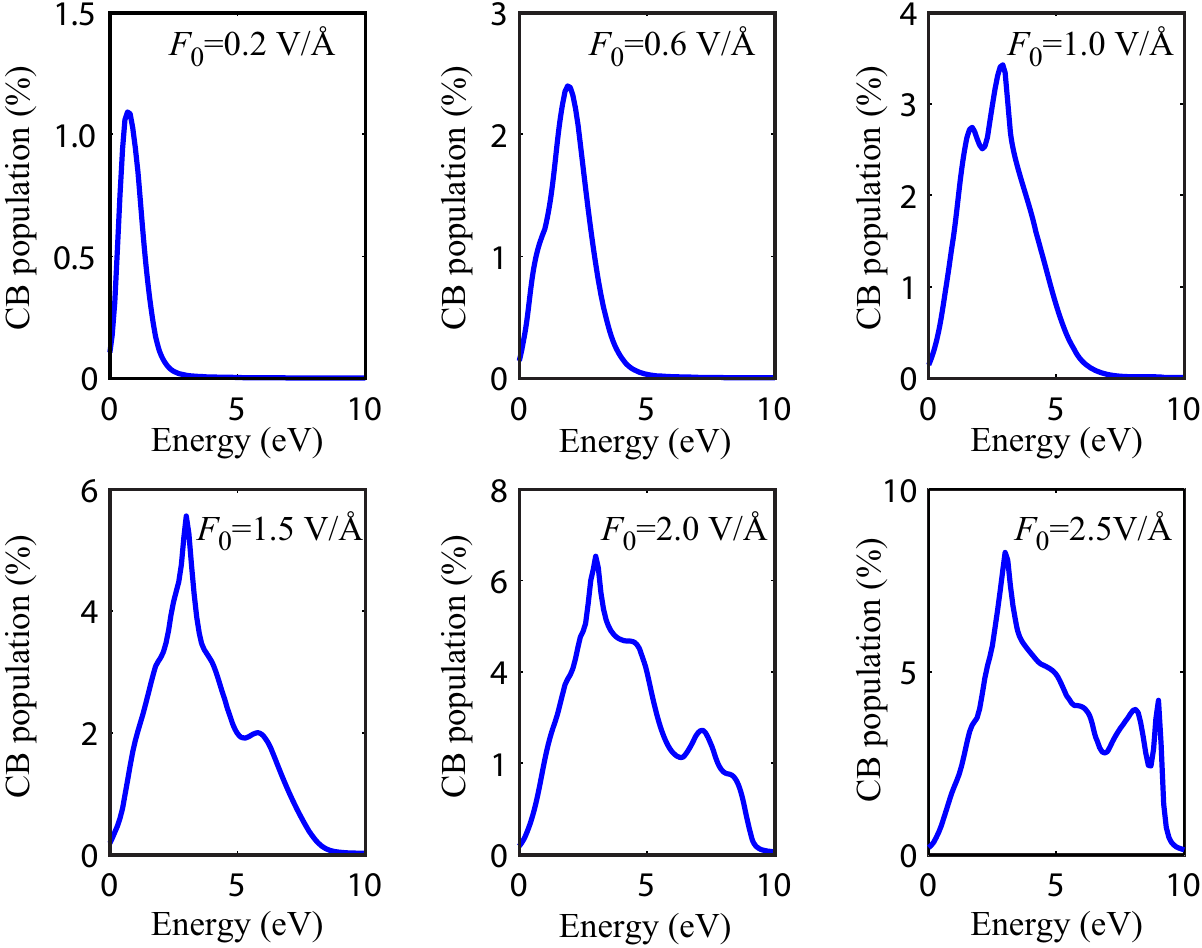}\end{center}
\caption{Residual conduction band population $N_{c,E}(E)$ is 
shown as a function of energy for different amplitudes of the optical pulse. The polarization of the optical pulse is along axis $x$.    
} 
\label{CBpopulation_Energy}
\end{figure}

Another characteristics of the electron redistribution between the conduction and valence bands 
is the residual conduction band population calculated as a function of electron energy. 
Such function is defined by the following expression 
\begin{equation}
{\cal N}_{c,E} (E) = \sum _{\vec{q}} |\beta _{c\vec{q}} (t\rightarrow \infty)|^2 
  \delta \left( E - E_c (\vec{q}) \right) ,
\label{NE}
\end{equation}
where $\delta$ is the Dirac $\delta$-function. The function ${\cal N}_{c,E} (E)$ is shown in 
Fig.\ \ref{CBpopulation_Energy} for different amplitudes $F_0$ of the optical pulse. The conduction band 
population as a function of energy has a single peak structure with well-defined maximum at 
finite electron energy. For example, for $F_0 = 0.6$ V/\AA, the maximum of ${\cal N}_{c,E} (E)$ is 
at $E \approx 2$ eV. The width of the peak  also increases with increasing the pulse amplitude. At 
$F_0 = 1.5$ V/\AA \ the peak occupies the whole conduction band, i.e. after the pulse ends all the conduction 
band states are partially occupied by electrons. 
The conduction band population exactly at the Dirac point, i.e. at zero energy, is small. Such behavior is correlated with the distribution of the conduction band population in the reciprocal space shown in Fig.\ \ref{Population_KxKy_field}.

\subsection{Transferred charge}

The generated time-dependent electric current and the net transferred charge through the graphene layer are another characteristics of interaction of the electron system of graphene with laser pulse. The net 
transferred charge through the graphene system, which is calculated from expression (\ref{Qtr}), is determined 
by the residual state of the system. Namely, the residual polarization of the electron system is equal to the 
transferred charge. Within a single band approximation, when the interband coupling is assumed to be zero, the electron dynamics is completely reversible. In this case the transferred charge through the system is zero, although the generated time-dependent current during the propagation of the pulse is non-zero. Therefore, the 
transferred charge is non-zero only due to finite interband coupling, which results in irreversible electron 
dynamics and finite residual polarization of the electron system. 

\begin{figure}
\begin{center}\includegraphics[width=0.4\textwidth]{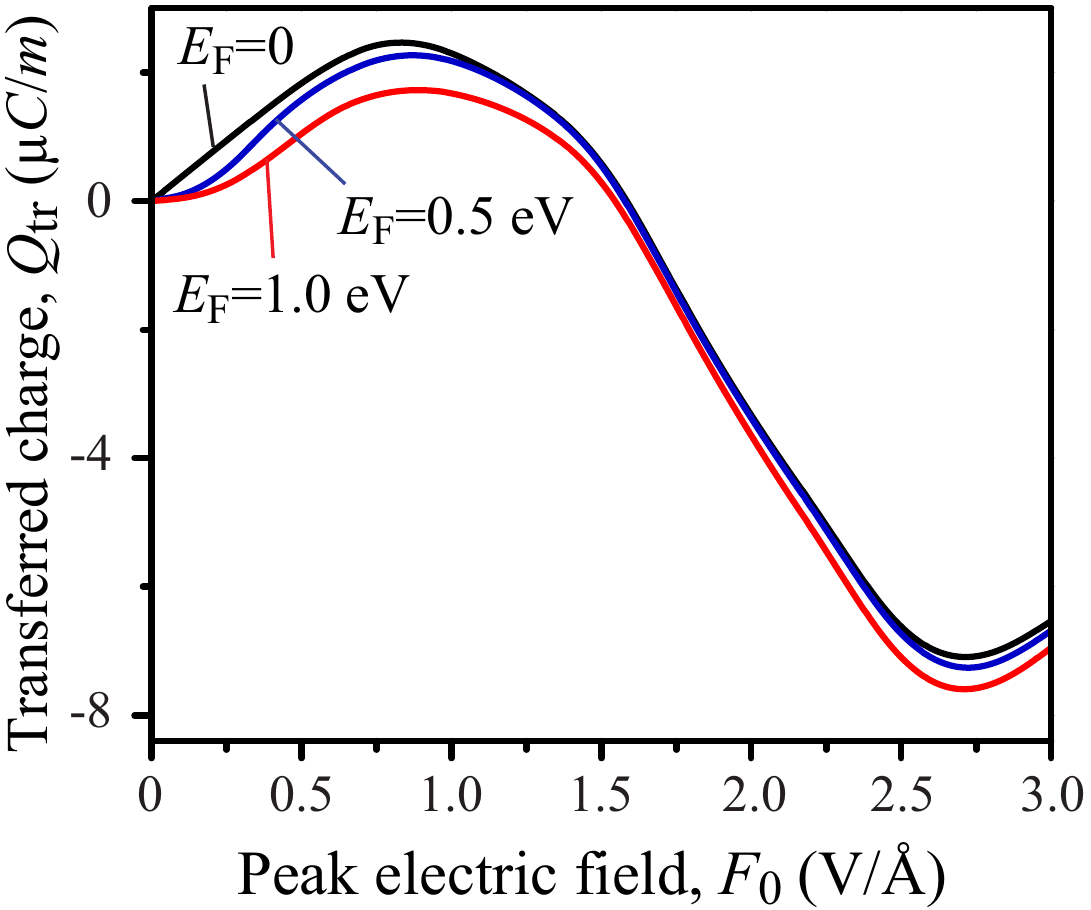}\end{center}
\caption{Transferred charge through graphene monolayer is shown as a 
function of the amplitude of the optical pulse $F_0$. The transferred 
charge is shown for different levels of doping of graphene, which 
is characterized by the electron Fermi energy $E_F$ 
in the conduction band. The polarization of the optical pulse is along 
axis $x$.  
} 
\label{Charge}
\end{figure}

In Fig.\ \ref{Charge} the transferred charge is shown as a function of the amplitude $F_0$ of the optical pulse. 
The results are shown for different levels of doping of the electron system. The level of the doping is characterized by the Fermi energy, $E_F$. The transferred charge has weak dependence on the doping. The data in Fig.\ \ref{Charge} are shown for polarization of the optical pulse along axis $x$, i.e. along the axis of  symmetry of the crystal structure of graphene. Such polarization corresponds to angle $\theta =0$ and  
the charge in this case is  transferred only along axis $x$. For other polarization of the optical pulse, $30^0 > \theta > 0^0$, there is non-zero transferred charge in the direction perpendicular to the polarization of the 
pulse, but it is very small compared to the  charge transferred along the direction of polarization. This 
property illustrates that the symmetry of the crystal structure of graphene is close to cylindrical. 

An interesting property of the transferred charge (see Fig.\ \ref{Charge}) is the change of its sign with increasing pulse amplitude. 
While at small pulse intensities, $F_0 \lesssim 1.5$ V/\AA, the transferred charge is positive, i.e. the charge is transferred in the direction of the 
pulse-field maximum, at large pulse intensities, $F_0> 1.5$ V/\AA, the 
transferred charge is negative, i.e. it is transferred in the direction 
opposite to the direction of the pulse maximum.
 This is the combination of dielectric and metal behaviors, where for dielectrics the transferred charge is 
positive, while for metal the transferred charge is negative. Therefore, in terms of the transferred charge the graphene monolayer behaves as a dielectrics at low pulse intensities and as a metal at large intensities.  
With increasing the dopping, i.e., with increasing the Fermi energy, $E_F$, 
the positive transferred charge at low intensities decreases, making the graphene system more metallic in terms of the charge transfer.

The origin of the change of the sign of the transferred charge can be understood 
from the expression for generated electric current. The generated current [see
Eq.\ (\ref{current})] has two contributions: intraband contribution, which is determined by intraband matrix elements of velocity operator, ${\cal V}^{cc}$ and 
${\cal V}^{vv}$, and interband contribution, which is proportional to the interband 
matrix elements of the velocity ${\cal V}^{cv}$. The interband contribution to the current is  almost an order of magnitude  smaller than the corresponding intraband contribution. Such large difference of these two contributions is due to 
their different dependence on the expansion coefficients  
$\beta _{\alpha \vec{q}} (t) $, where $\alpha =c$ or $v$. The intraband terms in the expression for the current are proportional to  
$\left[ \left| \beta _{c \vec{q}} (t) \right|^2 - \left| \beta _{v \vec{q}} (t) \right|^2 \right]$, while the intraband terms are determined by 
combination $\beta _{c \vec{q}}^{*} (t)\beta _{v \vec{q}} (t)$. The conduction
band population as a function of the wave vector has a structure of localized 
spots with large value  of $\beta _{c \vec{q}} (t) $ ($\beta _{c \vec{q}} (t)$ is almost 1 at these spots). At these spots, the terms $\left[ \left| \beta _{c \vec{q}} (t) \right|^2 - \left| \beta _{v \vec{q}} (t) \right|^2 \right]$, 
which determine the intraband current, is almost 1, while the factors $\beta _{c \vec{q}}^{*} (t)\beta _{v \vec{q}} (t)$ are almost 0, which results in 
small generated interband current. 

Taking into account only intraband contribution to the generated current, 
we rewrite Eq.\ (\ref{current}) for the $x$ component of the current 
in the following form
\begin{equation}
J_x (t) = \frac{e}{a^2} \sum_{\vec{q}}  
\left( 2 \left| \beta _{c \vec{q}} (t) \right|^2  -1 \right)
    {\cal V}_{x}^{cc}(\vec{k}_T(\vec{q},t)) ,
\label{currentIntra}
\end{equation}
where we took into account the electron-hole symmetry of the nearest neighbor 
model of graphene, i.e. ${\cal V}_{x}^{cc} = - {\cal V}_{x}^{vv}$, and used 
the property $\left| \beta _{c \vec{q}} (t) \right|^2 + \left| \beta _{v \vec{q}} (t) \right|^2 = 1$.

\begin{figure}
\begin{center}\includegraphics[width=0.4\textwidth]{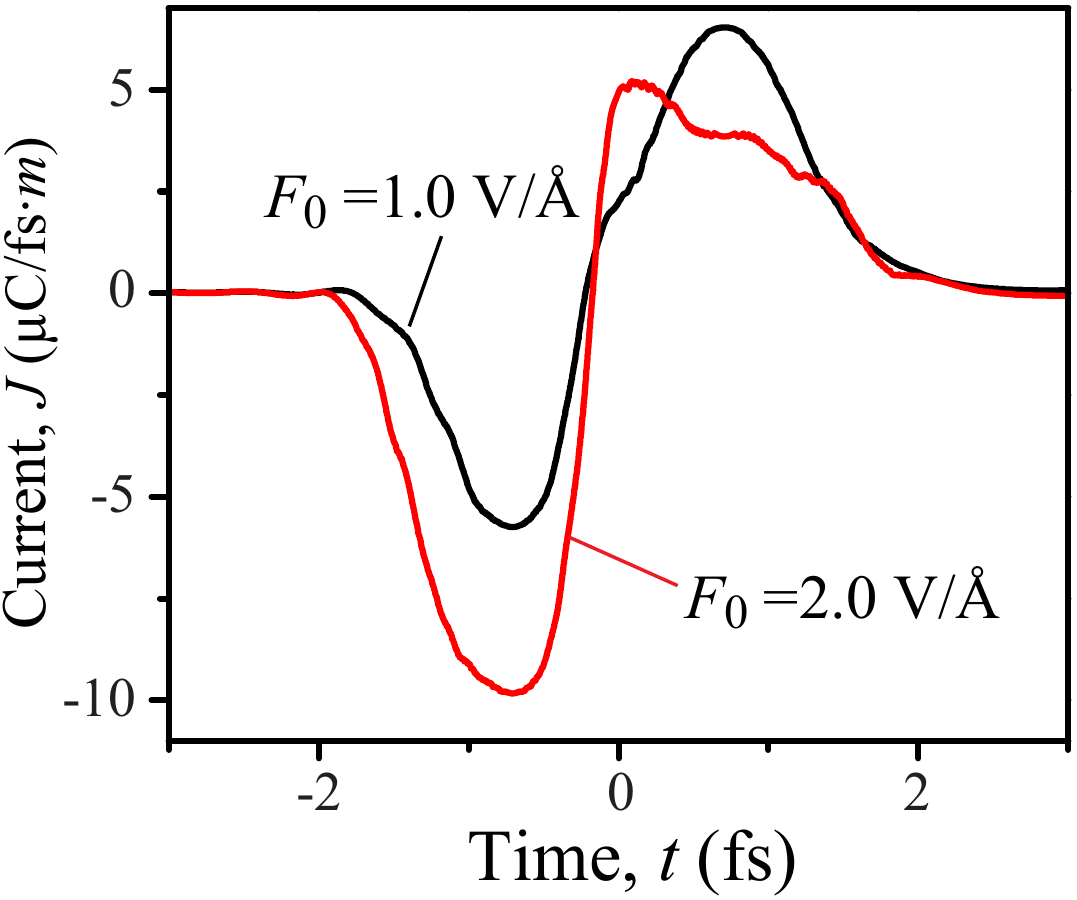}\end{center}
\caption{The electric current generated in graphene system by the 
electric field of the optical pulse is shown as a function of time for 
two amplitudes of the optical pulse: $F_0 = 1.0$ V/\AA \ and $F_0= 2.0$ V/\AA. The transferred charge in these two cases is positive for 
$F_0 = 1.0$ V/\AA \ and negative for $F_0 = 2.0$ V/\AA. 
The polarization of the optical pulse is along 
axis $x$.  
} 
\label{Fcurrent}
\end{figure}

At small pulse amplitudes, the main mechanism, which determines the transferred 
charge, is an increase of the conduction band population, i.e. $\beta _{c \vec{q}} (t)$, due interband electron dynamics. The corresponding 
generated current is  shown in Fig.\ \ref{Fcurrent} for $F_0 = 1.0$ V/\AA. At
$t<0$ the current is negative, while at $t>0$ the current is positive. Since the 
conduction band population at later moments of time ($t>0$) is larger 
than the conduction band population at $t<0$ [see Fig.\ \ref{CB_field}] then the 
positive component of the current is larger than the negative one, which results in positive transferred charge through the system. 

At large pulse amplitudes $F_0$, another mechanism becomes important for the
formation of the charge transfer. Namely, due to intraband electron dynamics, 
the matrix elements of the velocity operator should be calculated at the 
time-dependent wave vector $\vec{k}_T(\vec{q},t)$, which is proportional 
to the amplitude of the electric field. Since the velocity ${\cal V}_{x}^{cc}$ 
becomes small away from the Dirac points, then at large $F_0$, when  
$|\vec{k}_T(\vec{q},t) - \vec{q}|$ becomes large, there is an additional suppression of the current. The corresponding generated current is shown in 
Fig.\ \ref{Fcurrent} for $F_0 = 2.0$ V/\AA. The data clearly show suppression 
of the positive current at $t>0$, which results in negative transferred charge. 
This mechanism, which is due to intraband electron dynamics, of generating charge transfer, is similar to metals in strong optical pulse.\cite{Metal_PRB_2013}

Thus, for graphene, at low amplitudes of the optical pulse, the interband dynamics determines the transferred charge, which is positive in this case, which is similar to dielectrics. At high amplitudes of the pulse, the 
intraband dynamics provides the main contribution to the transferred charge, making it negative similar to metals.

\section{Conclusion} 

Interaction of ultrashort and strong optical pulse with graphene is determined by unique 
gapless energy dispersion law of electron system in graphene. Such energy dispersion results also 
in strong dependence of the interband dipole matrix elements on the wave vector in the reciprocal space
with singularity of the interband interaction at the Dirac points. As a result, the electron 
dynamics in the time-dependent electric field of the laser pulse is irreversible with large 
residual conduction band population. The residual conduction band population is also 
comparable with the maximum conduction band population realized during propagation of the pulse. 

Due to singularity of the interband dipole interaction, the residual conduction band population, i.e. the 
conduction band population after the pulse ends, shows strong dependence on the electron wave vector. In 
the reciprocal space the conduction band population shows a few highly localized spots, where the electrons 
are almost completely transferred from the valence band to the conduction band states. Such spots of high conduction band population are concentrated near the Dirac points and the number of the spots depends on the 
pulse intensity.

Another characteristics of interaction of the optical pulse with graphene is generated transferred charge through 
the system. 
Due to highly irreversible electron dynamics in the optical pulse, the transferred charge is nonzero and 
can be positive or negative depending on the pulse intensity. For the optical pulse with small intensity, i.e. 
with the amplitude $F_0 \lesssim 1.5$ V/\AA, the direction of the transferred charge is the same as the direction 
of the pulse maximum, which is similar to dielectrics. For the pulse with large intensity, $F_0> 1.5$ V/\AA, the charge is transferred in the direction opposite to the direction of the pulse maximum, which is the behavior expected in metals. This property can be used to control the direction of the transfer of the electric charge in graphene in optical pulse by varying the intensity of the pulse. In terms of the direction of the transferred charge, the graphene behaves as a metal or dielectrics depending on the pulse amplitude. 

\section*{Acknowledgment}

This work was supported by the Max Planck Society and
the Deutsche Forschungsgemeinschaft Cluster of Excellence:
Munich Center for Advanced Photonics (http://www.munich-photonics.de). Major funding was provided by Grant No. DE-FG02-01ER15213 from the Chemical Sciences, Biosciences
and Geosciences Division. Supplementary funding came from
Grant No. DE-FG02-11ER46789 from the Materials Sciences
and Engineering Division of the Office of the Basic Energy
Sciences, Office of Science, U.S. Department of Energy, and
Grant No. ECCS-1308473 from NSF.


\end{document}